\documentclass[a4paper,10pt,twoside]{cpc-hepnp}
\usepackage{CJK,upgreek,fancyhdr}
\usepackage{multicol}
\usepackage{graphicx}
\usepackage{booktabs}
\usepackage{amssymb,bm,mathrsfs,bbm,amscd}
\usepackage[tbtags]{amsmath}
\usepackage{lastpage}
\usepackage{color}

\begin{document}
\begin{CJK*}{GB}{gbsn}

\fancyhead[c]{\small Chinese Physics C~~~Vol. xx, No. x (2020) xxxxxx}


\title{The $ss \bar s \bar s$ tetraquark states and the observed structure $X(2239)$ by BESIII Collaboration\thanks{This project is supported by the National Natural Science Foundation of China under Grants No.~11705056, No.~11475192 and No.~U1832173. This work is also supported by the Sino-German CRC 110 "Symmetries and the Emergence of Structure in QCD" project by NSFC under the Grant No.~11621131001, and the Key Research Program of Frontier Sciences, CAS, under the Grant No.~Y7292610K1. }}

\author{%
      Qi-Fang L\"u (ÂÀÆë·Å)$^{1,2;1)}$\email{lvqifang@hunnu.edu.cn}%
\quad Kai-Lei Wang (Íõ¿­À×)$^{3;2)}$\email{wangkaileicz@foxmail.com}%
\quad Yu-Bing Dong (¶­Óî±ø)$^{4,5,6;3}$\email{dongyb@ihep.ac.cn}
}
\maketitle

\address{%
$^1$ Department of Physics, Hunan Normal University, and Key Laboratory of Low-Dimensional Quantum
Structures and Quantum Control of Ministry of Education, Changsha 410081, China\\
$^2$ Synergetic Innovation Center for Quantum Effects and Applications (SICQEA),
Hunan Normal University, Changsha 410081, China\\
$^3$ Department of Electronic Information and Physics, Changzhi University, Changzhi, Shanxi,046011,China\\
$^4$ Institute of High Energy Physics, Chinese Academy of Sciences, Beijing 100049, China\\
$^5$ Theoretical Physics Center for Science Facilities (TPCSF), CAS, Beijing 100049, China\\
$^6$ School of Physical Sciences, University of Chinese Academy of Sciences, Beijing 101408, China
}

\begin{abstract}
We investigate the mass spectrum of the $ss \bar s \bar s$ tetraquark states within the relativized quark model. By solving the Schr\"{o}dinger-like equation with the relativized potential, the masses of the $S-$ and $P-$wave $ss \bar s \bar s$ tetraquarks are obtained. The screening effects are also taken into account. It is found that the observed resonant structure $X(2239)$ in the $e^+e^- \to K^+K^-$ process by BESIII Collaboration can be assigned as a $P-$wave $1^{--}$ $ss \bar s \bar s$ tetraquark state. Furthermore, the radiative transition and strong decay behaviors of this structure are also estimated, which can provide helpful information for future experimental searches.
\end{abstract}

\begin{keyword}
Tetraquark, Spectrum, Radiative and strong decays, Relativized quark model
\end{keyword}

\begin{pacs}
12.39.Ki, 13.40.Hq, 14.40.Cs
\end{pacs}

\footnotetext[0]{\hspace*{-3mm}\raisebox{0.3ex}{$\scriptstyle\copyright$}2019
Chinese Physical Society and the Institute of High Energy Physics
of the Chinese Academy of Sciences and the Institute
of Modern Physics of the Chinese Academy of Sciences and IOP Publishing Ltd}%

\begin{multicols}{2}

\section{Introduction}

Recently, the BESIII Collaboration analyzed the cross section of the $e^+e^- \to K^+K^-$ process at center-of-mass energies varying from 2.00 to 3.08 GeV. A resonant structure was observed in the line shape, which has a mass of $2239.2 \pm 7.1 \pm 11.3~\rm{MeV}$ and a width of
$139.8 \pm 12.3 \pm 20.6~\rm{MeV}$~\cite{Ablikim:2018iyx}. Given its production process, the quantum number of this resonant structure can be assigned as
$J^{PC} = 1^{--}$.

From the Review of Particle Physics, there exist four $J^{PC}=1^{--}$ observed states around 2.2 GeV, such as $\phi(2170)$, $\rho(2150)$, $\omega(2205)$ and $\rho(2270)$~\cite{Tanabashi:2018oca}. The $\phi(2170)$ state with $I^G(J^{PC}) = 0^-(1^{--})$, labeled previously as $Y(2175)$, has been investigated within many theoretical interpretations, which include conventional $s\bar s$ state~\cite{Ding:2007pc,Shen:2009zze,Wang:2012wa,Afonin:2014nya,Pang:2019ttv}, hybrid state~\cite{Ding:2007pc,Ding:2006ya}, tetraquark state~\cite{Wang:2006ri,Chen:2008ej,Drenska:2008gr,Chen:2018kuu,Ke:2018evd,Cui:2019roq}, $\Lambda \bar \Lambda(^3S_1)$ bound state or hexaquark state~\cite{Zhao:2013ffn,Deng:2013aca,Abud:2009rk,Dong:2017rmg,Cao:2018kos}, and $\phi K \bar K$ resonance state~\cite{MartinezTorres:2008gy,GomezAvila:2007ru}. The $\rho(2150)$, $\omega(2205)$ and $\rho(2270)$ are also studied as conventional radial or orbital excited mesons in the consistent quark model~\cite{Ebert:2009ub,He:2013ttg,Chen:2017sjs}. As mentioned by BESIII Collaboration, the newly observed resonant structure, denoted as $X(2239)$ in present work, differs from the masses and widths of the $\phi(2170)$ and $\rho(2150)$, which seems to be a new resonance~\cite{Ablikim:2018iyx}. The $\omega(2205)$ and $\rho(2270)$ listed in the further states are both broad states, which can not be the same structure as this newly observed one~\cite{Tanabashi:2018oca}.

In the conventional quark model, several highly excited $1^{--}$ $\rho$, $\omega$, and $\phi$ states are predicted in this energy region, and their strong decay behaviors have been investigated in the quark pair creation model. Due to the large phase space, the predicted total decay widths of these states are rather broad, which suggests that the newly observed state with about $140$ MeV width may not be a conventional excited meson. More exotic interpretations, such as the $ss \bar s \bar s$ tetraquark state, are needed to be considered to clarify its nature. In the literature, the $P-$wave $ss \bar s \bar s$ system was mostly investigated by the QCD sum rule method~\cite{Wang:2006ri,Chen:2008ej,Chen:2018kuu,Cui:2019roq,Wang:2019nln} or the simple quark models~\cite{Drenska:2008gr,Ke:2018evd}, and their results are inconsistent with each other. Hence, it is essential to study this system in a more realistic potential model.

In this work, we firstly employ a relativized quark model to estimate the masses of $ss\bar s \bar s$ tetraquark states. The relativized quark model, proposed by Godfrey, Capstick, and Isgur, has been widely used to study the properties of the conventional hadrons and gives a unified description of the traditional hadron spectra~\cite{Godfrey:1985xj,Capstick:1986bm,Godfrey:1998pd,Godfrey:2015dia,Ferretti:2013faa, Ferretti:2013vua,Lu:2014zua,Godfrey:2015dva,Song:2015nia,Song:2015fha,Godfrey:2014fga}. Also, it has been extended to investigate various tetraquark systems, such as $Qq \bar Q \bar q$, $Qq \bar q \bar q$ and so on~\cite{Lu:2016cwr,Lu:2016zhe,Anwar:2018sol}. Moreover, the relativistic effects are involved in this model, which may be essential for the up, down and strange quarks. Therefore, it is suitable to deal with the $ss\bar s \bar s$ tetraquark states, where strange quarks and antiquarks are included. To calculate the tetraquark masses, we restrict our works in the diquark-antidiquark picture, which has been extensively discussed and employed in the literature~\cite{Lu:2016cwr,Lu:2016zhe,Anwar:2018sol,Lu:2017meb,Maiani:2004vq,Brodsky:2014xia,Anselmino:1992vg,Kiselev:2017eic,Richard:2018yrm,Lebed:2015tna,Lebed:2016yvr,Ebert:2005nc,Ebert:2008kb,Ebert:2008id,Hadizadeh:2015cvx}.
We first calculate the masses and wave functions of the axial-vector and vector $ss$ diquarks, and then obtain the mass spectra and diquark-antidiquark wave functions by solving the Schr\"{o}dinger-type equation between the diquark and antidiquark. The total wave functions can be expressed as the multiplication of the diquark, antidiquark, and diquark-antidiquark wave functions. The predicted mass of the lowest $1^{--}$  $ss\bar s \bar s$ tetraquark is 2227 MeV, which is consistent with the experimental data $2239.2 \pm 7.1 \pm 11.3~\rm{MeV}$ by BESIII Collaboration. It suggests that the newly observed resonant structure $X(2239)$ can be assigned as the lowest $J^{PC}=1^{--}$ $ss \bar s \bar s$ tetraquark state. Then, using the wave functions obtained from the relativized quark model and the electromagnetic transition operator, we estimate the radiative decays of the $ss\bar s \bar s$ tetraquarks. It is found that the radiative decay width for the lowest $1^{--}$ $ss \bar s \bar s$ tetraquark state is 27 keV, which is significant. Furthermore, the strong decay behaviors are also discussed, and the ratios of the dominating channels are estimated. The information on radiative and strong decays may be useful for future experimental searches.

This paper is organized as follows. In Sec. 2, the relativized quark model is briefly introduced, and the masses of $ss\bar s \bar s$ tetraquark states are calculated. In Sec. 3, the radiative transitions and strong decays of $ss\bar s \bar s$ tetraquark states are numerically estimated. Finally, we give a short summary in the last section.

\section{Mass spectrum}{\label{spectrum}}

The Hamiltonian between the quark and antiquark in the relativized quark model can be expressed as
\begin{equation}
\tilde{H} = H_0+\tilde{V}(\boldsymbol{p},\boldsymbol{r}), \label{ham}
\end{equation}
with
\begin{equation}
H_0 = (p^2+m_i^2)^{1/2}+(p^2+m_j^2)^{1/2},
\end{equation}
\begin{equation}
\tilde{V}(\boldsymbol{p},\boldsymbol{r}) = \tilde{H}^{conf}_{ij}+\tilde{H}^{cont}_{ij}+\tilde{H}^{ten}_{ij}+\tilde{H}^{so}_{ij},
\end{equation}
where the $\tilde{H}^{conf}_{ij}$ includes the spin-independent linear confinement and Coulomb-like interaction, the $\tilde{H}^{cont}_{ij}$, $\tilde{H}^{ten}_{ij}$, and $\tilde{H}^{so}_{ij}$ are the color contact term, the color tensor interaction, and the spin-orbit term, respectively. The $\tilde{H}$ represents that the operator $H$ has taken account of the relativistic effects via the relativized procedure. The explicit forms of these interactions and the details of this relativization scheme can be found in Ref.~\cite{Godfrey:1985xj,Capstick:1986bm}. In the original GI model, the coupled channel or screening effects are ignored, which may influence on the properties of the excited mesons and tetraquarks~\cite{Dong:1994zj,Ding:1995he,Song:2015nia,Lu:2016cwr,Pang:2017dlw,Pang:2018gcn}. The modified procedure $br \to b(1-e^{-\mu r})/\mu$  with a new screening parameter $\mu$ performs a better description of meson and tetraquark spectra, especially for the strange quark systems~\cite{Song:2015nia,Lu:2016cwr}. Hence, we take the screening effects into account for completeness.

In present work, only the antitriplet diquark $[\bar 3_c]_{ss}$ are considered, while the $[6_c]_{ss}$ type diquarks can not be formed in the GI quark model. For the quark-quark interaction in the antitriplet diquark and triplet antidiquark systems, the relation $\tilde{V}_{ss}(\boldsymbol{p},\boldsymbol{r}) = \tilde{V}_{\bar s \bar s}(\boldsymbol{p},\boldsymbol{r}) = \tilde{V}_{s\bar s}(\boldsymbol{p},\boldsymbol{r})/2$ is employed. The parameters used in our calculations are the same as the ones in the original work~\cite{Godfrey:1985xj}. The structures of the $ss\bar s \bar s$ tetraquarks are illustrated in Fig.~\ref{tetra}. The interaction between diquark and antidiquark $\tilde{V}_{{ss}-{\bar s \bar s}}(\boldsymbol{p},\boldsymbol{r})$ equals to the quark-antiquark interaction $\tilde{V}_{s\bar s}(\boldsymbol{p},\boldsymbol{r})$.  The ground $ss$ diquark lies in $S-$wave, and has the spin-parity $J^P=1^+$ named as axial-vector diquark. For the excited ones, we only consider the $P-$wave $ss$ diquark with spin-parity $J^P=1^-$, which is denoted as vector diquark.

\begin{center}
\includegraphics[width=5cm]{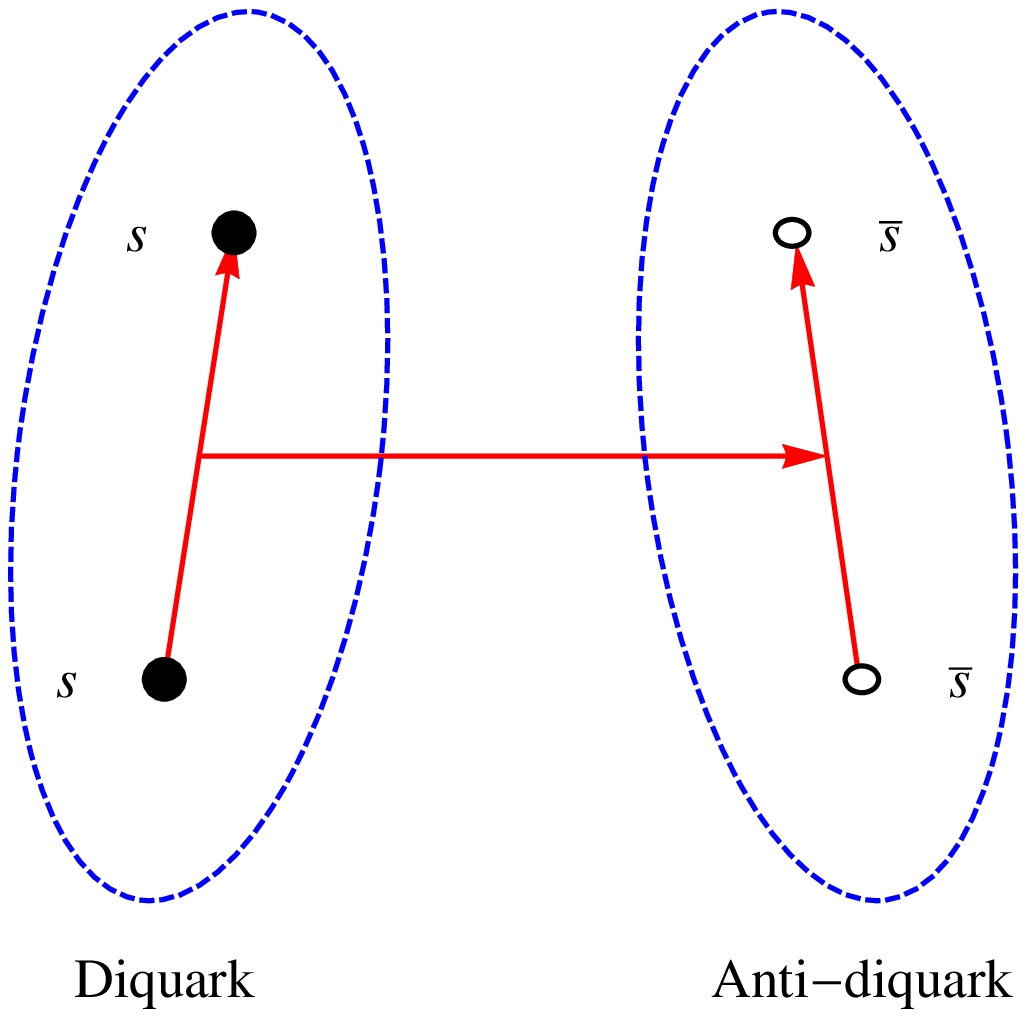}
\figcaption{\label{tetra}   The $ss\bar s \bar s$ tetraquark states with Jacobi coordinates. }
\end{center}

It should be mentioned that a constituent diquark naturally has a size as well as a constituent quark, though they are treated as point-like in the potential formula~\cite{Anselmino:1992vg}. The constituent quark model works whether the constituent quark or diquark is sizable or not. The comparison between diquark picture and full few-body calculation can be found in Ref.~\cite{Kiselev:2017eic,Richard:2018yrm}. Certainly, trying to investigate the $P-$wave tetraquarks with four-body calculation in relativized quark model is more interesting, convincing and complicated.

Here, we use the Gaussian expansion method to solve the Hamiltonian~(\ref{ham}) with $\tilde{V}_{ss}(\boldsymbol{p},\boldsymbol{r})$ potential~\cite{Hiyama:2003cu}. The obtained masses of the axial-vector and vector $ss$ diquarks are presented in Table.~\ref{tab1}. Since the $\mu=0.02~\rm{GeV}$ case can give a rather better description of the strange quark systems~\cite{Song:2015nia,Lu:2016cwr}, we would like to use the diquark masses at this value to calculate the masses and wave functions of $ss\bar s \bar s$ tetraquarks.

\end{multicols}
\begin{center}
\tabcaption{ \label{tab1} Obtained masses of the axial-vector and vector $ss$ diquarks. $A$ and $V$ denote the axial-vector and vector diquarks, respectively. The brace corresponds to symmetric quark content in flavor. The units are in MeV.}
\footnotesize
\begin{tabular*}{170mm}{c@{\extracolsep{\fill}}cccc}
\toprule  Quark content &  Diquark type  & Mass (GI model)   & Mass $(\mu= 0.02~\rm{GeV})$ & Mass $(\mu= 0.04~\rm{GeV})$ \\\hline
 $\{s,s\}$     &  $A$            & 1.135       & 1.121              & 1.108                 \\
 $\{s,s\}$     &  $V$            & 1.424       & 1.396              & 1.369                 \\
\bottomrule
\end{tabular*}
\vspace{0mm}
\end{center}
\vspace{0mm}
\begin{multicols}{2}
With the diquarks listed in Table.~\ref{tab1}, one can calculate the masses of the $ss\bar s \bar s$ tetraquarks and the wave functions between diquarks and anti-diquarks. Then, the total wave function of the $ss\bar s \bar s$ tetraquark can be expressed as the multiplication of the diquark, anti-diquark and relative wave functions.  The masses of $ss\bar s \bar s$ tetraquark states composed of the $A$ and $V$ diquarks and antidiquarks are presented in Tab.~\ref{mass1} and Fig.~\ref{ssss}. The predicted mass of the lowest $0^{++}$ state is 1716 MeV, which is consistent with the $f_0(1710)$ state. For the $1^{+-}$ $ss\bar s \bar s$ state, only $h_1(1965)$ state listed in the PDG book lies in this energy region~\cite{Tanabashi:2018oca}. Since the $h_1(1965)$ was observed in $\omega \eta$ and $\omega \pi \pi$ final states, which disfavors its assignment as $1^{+-}$ $ss\bar s \bar s$ tetraquark.  In Ref.~\cite{Cui:2019roq,Wang:2019nln}, the authors suggest that the new structure $X(2063)$ observed in the $J/\psi \to \phi \eta \eta^\prime$ by BESIII Collaboration~\cite{Ablikim:2018xuz} is a $1^{+-}$ $ss\bar s \bar s$ tetraquark candidate within the QCD sum rule method. However, our calculated mass is 100 MeV lower than the experimental mass, which does not support this interpretation. Considering the mass, spin parity, and $\phi \phi$ decay mode of the $f_2(2300)$, we may assign it as a $2^{++}$ $ss\bar s \bar s$ tetraquark state.

\end{multicols}
\begin{center}
\tabcaption{ \label{mass1} Masses of the $ss\bar s \bar s$ tetraquark states composed of the $A$ and $V$ diquarks and antidiquarks. For the $V$ diquark and $A$ antidiquark case, the linear combinations together with $V$ diquark and $A$ antidiquark are understood to form the eigenstates of charge conjugation~\cite{Lebed:2016yvr,Lu:2016cwr}. The units are in MeV.}
\footnotesize
\begin{tabular*}{170mm}{c@{\extracolsep{\fill}}cccccc}
\toprule $J^{PC}$                &  Diquark        & Anti-diquark           & $S$     & $L$      & Mass   & Candidate \\\hline
 $|0^{++}\rangle$      &  $A$              & $\bar A$                   & 0     & 0                &  1716                 &  $f_0(1710)$\\
 $|1^{+-}\rangle$      &  $A$              & $\bar A$                  & 1     & 0                 &  1960                 &  \\
 $|2^{++}\rangle$        &  $A$              & $\bar A$                   & 2     & 0            &  2255                  &  $f_2(2300)$\\\hline
 $|0^{-\pm}\rangle$       & $V$           & $\bar A$                    & 0     & 0                 & 2004                &  \\
 $|1^{-\pm}\rangle$       & $V$           & $\bar A$                     & 1     & 0              & 2227                  &  $X(2239)$\\
 $|2^{-\pm}\rangle$       & $V$           & $\bar A$                     & 2     & 0                & 2497               &  \\\hline
 $|0^{-+}\rangle$      &  $A$              & $\bar A$                   & 1     & 1               &  2450                & $X(2500)$ \\
 $|1^{-+}\rangle$      &  $A$              & $\bar A$                    & 1     & 1               &  2581              &  \\
 $|1^{--}\rangle$      &  $A$              & $\bar A$                    & 0     & 1                &  2574             &  \\
 $|1^{--}\rangle$      &  $A$              & $\bar A$                    & 2     & 1               &  2468               &  \\
 $|2^{-+}\rangle$      &  $A$              & $\bar A$                   & 1     & 1               &  2619               &  \\
 $|2^{--}\rangle$      &  $A$              & $\bar A$                  & 2     & 1                &  2622               &  \\
 $|3^{--}\rangle$      &  $A$              & $\bar A$                  & 2     & 1               &  2660                &  \\
\bottomrule
\end{tabular*}
\vspace{0mm}
\end{center}
\vspace{0mm}
\begin{multicols}{2}

For the $P-$wave $ss\bar s \bar s$ tetraquarks, we predict three $1^{--}$ states. The lowest one has the internal excitation in the diquark or antidiquark, while the others have the relative excitations between diquarks and antidiquarks. The three $1^{--}$ states together with other theoretical works are listed in Tab.~\ref{mass2} for comparisons. It can be seen that our quark model classification is significantly different with the QCD sum rule works~\cite{Wang:2006ri,Chen:2008ej,Chen:2018kuu,Wang:2019nln}, and the authors did not consider the internal excitation of the diquark or anti-diquark within the simple quark model~\cite{Drenska:2008gr}. The predicted lowest one has a mass of 2227 MeV, which agrees well with the $X(2239)$ observed by BESIII Collaboration~\cite{Ablikim:2018iyx}. The experimental mass of the $\phi(2170)$ is about 50 MeV lower than our calculation, which can not be excluded as the $ss\bar s \bar s$ tetraquarks. The evidences of these other two higher $1^{--}$ states may have been obsevered in the previous experiments~\cite{Chen:2018kuu,Aubert:2007ur,Ablikim:2007ab,Shen:2009zze,Ablikim:2014pfc}, or are waiting to be discovered in future searches.

Furthermore, we predict several higher $ss\bar s \bar s$ tetraquarks around 2.5 GeV. For the higher $0^{-+}$ state, there exists a candidate $X(2500)$ with mass of $2470^{+15+101}_{-19-23}~\rm{MeV}$ observed in the $J/\psi \to \gamma \phi \phi$ process by BESIII Collaboration~\cite{Ablikim:2016hlu}. In the conventional quark model, the $X(2500)$ was assigned as the $\phi(5^1S_0)$ state given its mass and total width~\cite{Pan:2016bac,Wang:2017iai} , but with a tiny $\phi \phi$ partial decay width. The $ss\bar s \bar s$ tetraquark interpretation of the $X(2500)$ may avoid this defect due to its falling apart mechanism into the $\phi \phi$ final state. Other predictions can provide helpful information for future experimental searches.

\end{multicols}
\begin{center}
\tabcaption{ \label{mass2} The predicted three $1^{--}$ $ss\bar s \bar s$ tetraquarks states together with other theoretical works. The units are in MeV.}
\footnotesize
\begin{tabular*}{170mm}{c@{\extracolsep{\fill}}cccccc}
\toprule $J^{PC}$ &    Ours  & SQM~\cite{Drenska:2008gr} & QCDSR~\cite{Wang:2006ri}  & QCDSR~\cite{Chen:2008ej} & QCDSR~\cite{Chen:2018kuu} &  QSR~\cite{Wang:2019nln}      \\\hline
 $1^{--}$     &  2227       &                        & $2210\pm90$                  & $2300\pm400$             & $2340\pm170$              &  $3080\pm110$                \\
 $1^{--}$     &  2468       & 2243                   &                              &                          & $2410\pm250$              &       \\
 $1^{--}$     &  2574       & 2333                  &                               &                          &                           &         \\
\bottomrule
\end{tabular*}
\vspace{0mm}
\end{center}
\vspace{0mm}
\begin{multicols}{2}

\begin{center}
\includegraphics[width=8cm]{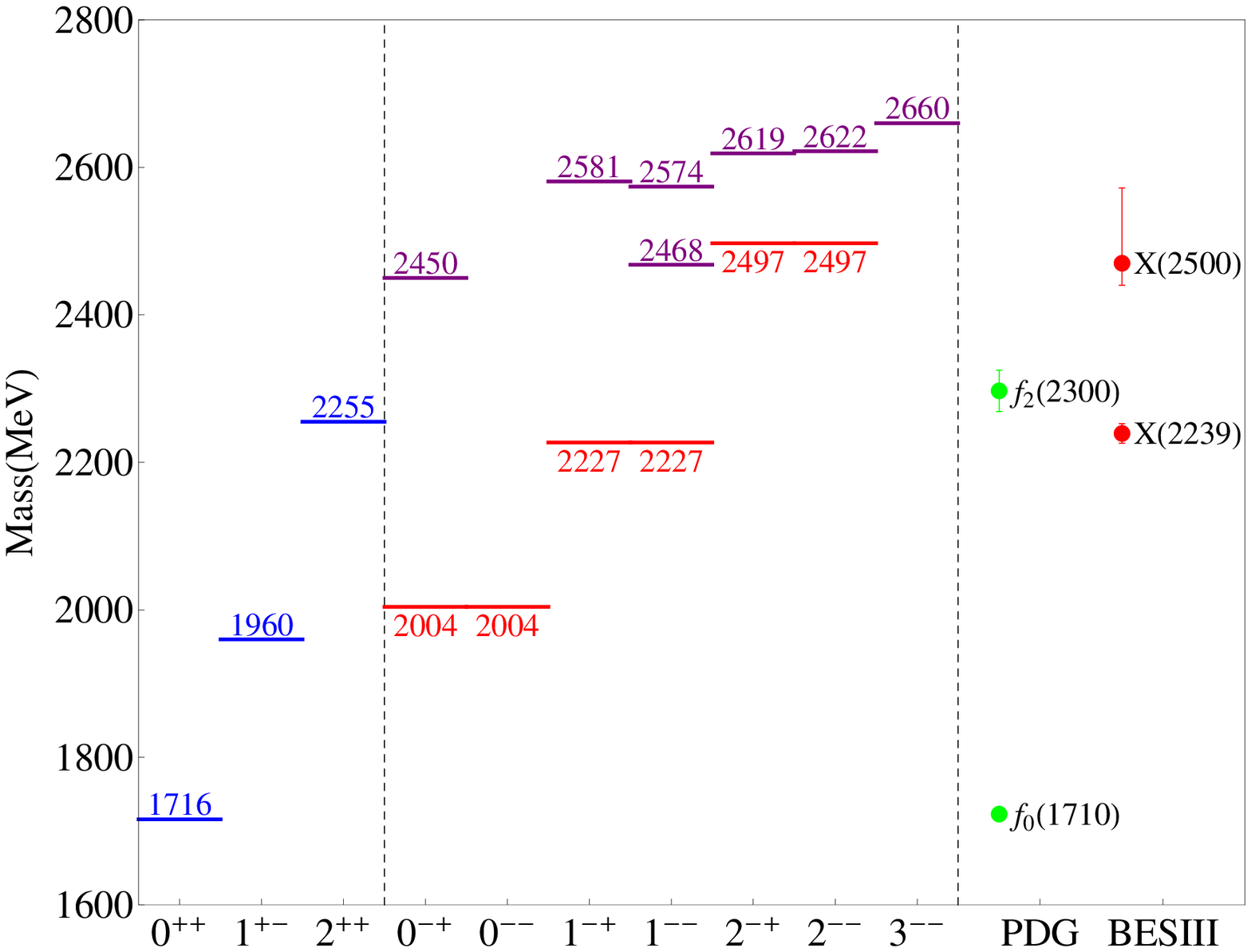}
\figcaption{\label{ssss}   Masses of the $ss\bar s \bar s$ tetraquark states. }
\end{center}

\section{Decays}{\label{Decays}}

\subsection{Radiative transitions}

Besides the mass spectrum, the decay behaviors are also needed to clarify these tetraquark states in future experiments. We firstly calculate the radiative transitions and then estimate the relevant ratios of the dominating strong decay modes.
To treat the radiative transitions between these $ss\bar s \bar s$ tetraquarks, one can adopt
an EM transition operator which has been successfully applied to study
the radiative decays of quarkonium and baryons~\cite{Deng:2016stx,Deng:2016ktl,Wang:2017hej}.
In this model, the quark-photon EM coupling at the tree level
is taken as
\begin{equation}
H_e = -\sum_j e_j \bar \psi_j \gamma_\mu^j A^\mu(\boldsymbol{k},\boldsymbol{r}_j) \psi_j, \label{he1}
\end{equation}
where $\psi_j$ stands for the $j$th quark field with coordinate $\boldsymbol{r}_j$ and $A^\mu$ is the photon field with three-momentum $\boldsymbol{k}$. To match the wave functions obtained by the Schr\"{o}dinger-like equation,
we adopt this quark-photon EM coupling in a nonrelativistic form.  In the initial-hadron-rest system,
the approximate form can be written as~\cite{Deng:2016stx,Deng:2016ktl,Wang:2017hej,Brodsky:1968ea,Li:1997gd,Zhao:2002id,Xiao:2015gra,Zhong:2011ti,Zhong:2011ht}
\begin{equation}
h_e \cong \sum_j \Bigg [ e_j \boldsymbol{r}_j \cdot \boldsymbol \epsilon - \frac{e_j}{2m_j} \boldsymbol \sigma_j \cdot (\boldsymbol \epsilon \times \boldsymbol {\hat k}) \Bigg ] e^{-i\boldsymbol{k} \cdot \boldsymbol{r}_j}, \label{he2}
\end{equation}
where $e_j$, $m_j$, and $\boldsymbol \sigma_j$ stand for the charge, consistent mass, Pauli spin vector for the $j$th quark, respectively. The $\boldsymbol \epsilon$ is the polarization vector of the final photon.

One can obtain the helicity amplitude $\mathcal{A}$ of the radiative transition~\cite{Deng:2016stx,Deng:2016ktl}
\begin{equation}
\mathcal{A} = -i \sqrt{\frac{\omega_\gamma}{2}} \langle f | h_e | i \rangle.
\end{equation}
Then, we can estimate the radiative transitions straightforward~\cite{Deng:2016stx,Deng:2016ktl}
\begin{equation}
\Gamma = \frac{|\boldsymbol k|^2}{\pi} \frac{2}{2J_i+1} \frac{M_f}{M_i} \sum_{J_{fz},J_{iz}} |\mathcal{A}|^2,
\end{equation}
where $J_i$ is the total angular momentum of the initial tetraquarks, and $J_{fz}$ and $J_{fi}$ are the components of the total angular momenta along the $z$ axis of the initial and final tetraquarks, respectively. In present work, the masses and wave functions of the $ss\bar s \bar s$ tetraquarks are adopted from our theoretical predictions.

The radiative transitions of the $ss\bar s \bar s$ tetraquarks are estimated and listed in Tab.~\ref{rad}. Here, we eliminate the notation $AA$ of the three ground states without causing misunderstanding. The predicted radiative transitions of the three ground states $0^{++}$, $1^{+-}$, and $2^{++}$ are
\begin{eqnarray}
\Gamma[|1^{+-} \rangle \to |0^{++} \rangle\gamma] = 157~\rm{keV},\\
\Gamma[|2^{++} \rangle \to |1^{+-} \rangle\gamma] = 175~\rm{keV},
\end{eqnarray}
respectively, which are significant large. As we assign the $f_0(1710)$ and $f_2(2300)$ as the $0^{++}$ and $2^{++}$ states respectively, the rather large radiative decay rates are useful to search for the missing $1^{+-}$ $ss\bar s \bar s$ tetraquark. Since the $0^{++}$ state has large branching ratios of $K\bar K$ and $\eta \eta$, more studies of the $ ss\bar s \bar s(1^{+-}) \to ss\bar s \bar s(0^{++}) \gamma \to K\bar K \gamma $ and $ ss\bar s \bar s(1^{+-}) \to ss\bar s \bar s(0^{++}) \gamma \to \eta \eta \gamma$ decay processes are suggested in future experiments.

For the transitions between $V\bar A$ type and ground states, the partial radiative decay widths range from 1 eV to tens keV. The $|V\bar{A}, 1^{--} \rangle \to |0^{++} \rangle\gamma$ process is 26.6 keV, which shows that the newly observed $X(2239)$ state has a significant radiative decay width. The other two $1^{--}$ states with relative excitations between diquarks and anti-diquarks can decay into $0^{++}$ and $2^{++}$ ground states, respectively,
\begin{eqnarray}
\Gamma[|A\bar{A}, 1^{--} \rangle_{S=0} \to |0^{++} \rangle\gamma] = 1137~\rm{keV},\\
\Gamma[|A\bar{A}, 1^{--} \rangle_{S=2} \to |2^{++} \rangle\gamma] = 119~\rm{keV},
\end{eqnarray}
where the ten times divergence of the partial widths derives from the different phase spaces. The radiative decay of the $S=0$ state to $2^{++}$ final state is highly suppressed, and also for the $S=2$ state to $0^{++}$ final state. These predictions may be helpful for searching and distinguishing the two higher $1^{--}$ $ss\bar s \bar s$ tetraquark states. About these radiative transitions of the excited $ss\bar s \bar s$ tetraquarks,
few discussions are found in the literature, thus, more theoretical and experimental studies are expected to be carried out in future.

\begin{center}
\tabcaption{ \label{rad}Radiative Transitions.}
\footnotesize
\begin{tabular*}{85mm}{c@{\extracolsep{\fill}}ccc}
\toprule Decay mode  & $M_i$(MeV)  & $M_f$(MeV) &Width(keV)  \\\hline

$|1^{+-} \rangle \to |0^{++} \rangle\gamma$   &1960   &1716        &157                       \\
$|2^{++} \rangle \to |1^{+-} \rangle\gamma$   &2255   &1960        &175                      \\\hline
$|V\bar{A}, 0^{-+} \rangle \to |1^{+-} \rangle\gamma$   &2004   &1960       &0.001    \\
$|V\bar{A}, 1^{--} \rangle \to |0^{++} \rangle\gamma$   &2227   &1716       &26.6                       \\
$|V\bar{A}, 1^{-+} \rangle \to |1^{+-} \rangle\gamma$   &2227   &1960       &3.1                    \\
$|V\bar{A}, 2^{--} \rangle \to |0^{++} \rangle\gamma$   &2497   &1716       &49.4                       \\
$|V\bar{A}, 2^{--} \rangle \to |2^{++} \rangle\gamma$   &2497   &2255       &4.6                         \\
$|V\bar{A}, 2^{-+} \rangle \to |1^{+-} \rangle\gamma$   &2497   &1960       &65.4                         \\\hline

$|A\bar{A}, 0^{-+} \rangle \to |1^{+-} \rangle\gamma$   &2450   &1960       &1345     \\
$|A\bar{A}, 1^{-+} \rangle \to |1^{+-} \rangle\gamma$   &2581   &1960       &1444                    \\
$|A\bar{A}, 1^{--} \rangle_{S=0} \to |0^{++} \rangle\gamma$   &2574   &1716       &1137                       \\
$|A\bar{A}, 1^{--} \rangle_{S=0} \to |2^{++} \rangle\gamma$   &2574   &2255       &0.0                       \\

$|A\bar{A}, 1^{--} \rangle_{S=2} \to |0^{++} \rangle\gamma$   &2468   &1716       &0.0                       \\
$|A\bar{A}, 1^{--} \rangle_{S=2} \to |2^{++} \rangle\gamma$   &2468   &2255       &119                       \\

$|A\bar{A}, 2^{-+} \rangle \to |1^{+-} \rangle\gamma$   &2619   &1960       &954                    \\

$|A\bar{A}, 2^{--} \rangle \to |2^{++} \rangle\gamma$   &2622   &2255       &809                       \\

$|A\bar{A}, 3^{--} \rangle \to |2^{++} \rangle\gamma$   &2660   &2255       &606                       \\

\bottomrule
\end{tabular*}
\vspace{0mm}
\end{center}
\vspace{0mm}

\subsection{Strong decays}

The strong decays can occur if the tetraquarks lie above the meson-meson or baryon-antibaryon threshold via falling apart mechanism. With the assignments of the $Y(2175)$ as the $ss\bar s \bar s$ tetraquark state, the authors estimate the ratios of the dominating decay modes by using this mechanism~\cite{Drenska:2008gr,Ke:2018evd}. In present work, three $J^{PC}=1^{--}$ $ss\bar s \bar s$ tetraquarks are predicted, and the ratios of some significant decay channels are studied as follows.

Apparently, the $S-$wave decay channels between the two final states are more favored than the $P-$wave channels. Also, the final states with strangeonium seem to be easier than the two kaon states where the initial two strange quarks should annihilate and two up/down quarks create simultaneously. For the baryon-antibaryon final states $\Lambda \bar \Lambda$ and $\Sigma \bar \Sigma$, more quark pairs should annihilate and create. Also, these processes are limited by the phase space, which are not considered in present work.

For the $S-$wave channels with strangeonium, three possible combinations of final states exist: $1^{--}$ and $0^{++}$, $1^{+-}$ and $0^{-+}$, and $1^{++}$ and $1^{--}$. Then, seven resonances with the $s\bar s$ component, $\eta$, $\eta^\prime$, $\omega$, $\phi$, $f_0(980)$, $h_1(1380)$, and $f_1(1420)$, should be considered. Because of the large mass of $f_1(1420)$, the $1^{++}$ and $1^{--}$ combinations are forbidden or highly limited by the phase space, which are neglected here. Finally, the possible $S-$wave decay channels with strangeonium are $\eta h_1(1380)$, $\eta^\prime h_1(1380)$, $f_0(980) \omega$, and $f_0(980)\phi$, where the nonet mixing angles $\theta_P = -11.3^\circ$ and $\theta_V = 39.2^\circ$ are adopted to determine the $s\bar s$ components in relevant mesons~\cite{Tanabashi:2018oca}. Following the route of Refs.~\cite{Drenska:2008gr,Ke:2018evd}, our results are listed in Tab.~\ref{str}. It is shown that the $f_0(980) \phi$ is the dominating decay mode, and the contribution of $\eta h_1(1380)$ is significant. Meanwhile, the $\eta^\prime h_1(1380)$ may be also important for the two higher states, while the ratios of the $f_0(980) \omega$ channel are rather small.

\end{multicols}
\begin{center}
\tabcaption{ \label{str} The ratios of strong decay channels of the predicted three $1^{--}$ $ss\bar s \bar s$ tetraquarks states. The short dash denotes the forbidden channel due to the phase space.}
\footnotesize
\begin{tabular*}{180mm}{c@{\extracolsep{\fill}}cccc}
\toprule $J^{PC}$ &   Mass(MeV)  &  $\Gamma(\eta h_1)$ : $\Gamma(\eta^\prime h_1)$ : $\Gamma(f_0(980) \omega)$ : $\Gamma( f_0(980) \phi)$ &  $\Gamma(\eta \omega)$ : $\Gamma(\eta^\prime \omega)$ : $\Gamma(\eta \phi)$ : $\Gamma(\eta^\prime \phi)$  &  $\Gamma(K K_1(1270))$ : $\Gamma(K K_1(1400))$ \\\hline
 $1^{--}$     &  2227       &   0.48~:~$--$~:~0.01~:~1                     & 0.01~:~0.01~:~1.35~:~1                  &  1.19~:~1    \\
 $1^{--}$     &  2468       &  0.69~:~0.37~:~0.01~:~1                  &   0.01~:~0.01~:~1.64~:~1                           & 1.54~:~1                     \\
  $1^{--}$     &  2574       & 0.78~:~0.54~:~0.01~:~1                 &    0.01~:~0.01~:~1.76~:~1                           & 1.68~:~1                      \\
\bottomrule
\end{tabular*}
\vspace{0mm}
\end{center}
\vspace{0mm}
\begin{multicols}{2}

Although the $P-$wave decay channels with strangeonium should contribute smaller than the $S-$wave modes, we also estimate the ratios of them. From Tab.~\ref{str}, it can be seen that the $\eta \phi$ and $\eta^\prime \phi$ final states may have significant contributions, while the $\eta \omega$ and $\eta^\prime \omega$ channels can be neglected.

For the $S-$wave channels with kaon states, only the $1^+$ plus $0^-$ channels $K K_1(1270)$ and $K K_1(1400)$ may play significant roles. The $1^-$ plus $0^+$ and $1^+$ plus $1^-$ combinations are forbidden or highly suppressed by the phase space. Our results show that the $K K_1(1270)$ and $K K_1(1400)$ channels have the comparable decay widths for these tetraquark states.

To sum up, the dominating channels of three $1^{--}$ tetraquark states are $f_0(980) \phi$ and $\eta h_1(1380)$, and the $\eta \phi$ and $\eta^\prime \phi$ modes may be also important. For the kaon final states, both $K K_1(1270)$ and $K K_1(1400)$ channels are important. We hope that the future experiments can search for the $1^{--}$ $ss\bar s \bar s$ tetraquark states in these channels.

\bigskip

\section{Summary}{\label{Summary}}
In this work, we investigate the masses of $ss \bar s \bar s$ tetraquark states within the relativized quark model proposed by Godfrey and Isgur. Here, only the antitriplet diquark $[\bar 3_c]_{cs}$ is considered. The masses of $ss \bar s \bar s$ tetraquark states are obtained by solving the Schr\"{o}dinger-like equation between diquark and antidiquark. The color screening effects are also added in present calculations. It is found that the newly observed resonant structure $X(2239)$ in the $e^+e^- \to K^+K^-$ process by BESIII Collaboration can be assigned as a $P-$wave $1^{--}$ $ss \bar s \bar s$ tetraquark state.

Besides the mass spectrum, the wave functions of the $ss \bar s \bar s$ tetraquark states are obtained simultaneously. Then, the radiative transitions between these tetraquarks and the ratios of the strong decay channels are estimated. The lowest $P-$wave $1^{--}$ $ss \bar s \bar s$ tetraquark state radiate to the ground state is 27 keV, and the mainly strong decay modes are $f_0(980) \phi$ and $\eta h_1(1380)$ final states. Moreover, other $ss \bar s \bar s$ tetraquark candidates $f_0(1710)$, $f_2(2300)$, and $X(2500)$ are also discussed here. We hope our assignments can be tested by future experiments.

\acknowledgments{We would like to thank Wen-Biao Yan, Xian-Hui Zhong and Dian-Yong Chen for valuable discussions. }
\end{multicols}

\vspace{-1mm}
\centerline{\rule{80mm}{0.1pt}}
\vspace{2mm}

\begin{multicols}{2}

\end{multicols}

\clearpage

\end{CJK*}
\end{document}